\documentclass{aa}  
\usepackage{graphicx}
\usepackage{natbib}
\bibpunct{(}{)}{;}{a}{}{,} 
\usepackage{txfonts}
\begin{document}
   \title{Interacting Jets from Binary Protostars}

   \author{G.C. Murphy
          \inst{1}
          \and
          T. Lery\inst{2}
          \and
          S. O'Sullivan\inst{3}
          \and
          D. Spicer\inst{4}
          \and
          F. Bacciotti\inst{5}
          \and
          A. Rosen\inst{6}
          }

   \offprints{G.C. Murphy}

   \institute{Laboratoire d'Astrophysique de Grenoble, CNRS,
				 Universit\'e Joseph Fourier, B.P. 53, F-38041 Grenoble, France\\
				 \email{Gareth.Murphy@obs.ujf-grenoble.fr}
				 \and
				 European Science Foundation, 1 quai Lezay-Marnesia, BP 90015,
				 67080 Strasbourg, France\\
				 \email{tlery@esf.org}
				 \and
				 UCD School of Mathematical Sciences,
				 University College Dublin, Belfield, Dublin
				 4, Ireland\\
				 \email{stephen.osullivan@ucd.ie}
				 \and
				 NASA/GSFC Laboratory for Solar and Space Physics, Mail Stop 612.1, Greenbelt,
				 MD 20771,
				 USA\\
				 \email{daniel.s.spicer@nasa.gov}
				 \and
				 INAF - Osservatorio Astrofisico di Arcetri, Largo E.
				 Fermi 5, 50125
				 Florence, Italy\\
				 \email{fran@arcetri.astro.it}
				 \and
				 Max-Planck-Institut f\"ur Radioastronomie, Auf dem
				 H\"ugel 69, D-53121 Bonn\\
				 \email{arosen@mpifr-bonn.mpg.de}
             }

   \date{Received xxx; accepted xxx}

 
  \abstract
   {}
   { We investigate potential models that could explain why multiple
		proto-stellar systems 
		predominantly show single jets.
		During their formation, stars most frequently produce energetic outflows
		and jets.
		However, binary jets have only been observed in a very small number of
		systems.}
   { We model numerically 3D binary jets for various outflow parameters. 
		We also model the propagation of jets from a specific source, namely L1551
		IRS
		5, known to have two jets, 
		using recent observations as constraints for simulations with a new MHD
		code. 
		We examine their morphology and dynamics, and produce synthetic emission
		maps.  }
   { We find that the two jets interfere up to the stage where
		one of them is almost destroyed or engulfed into the second one.
		We are able to reproduce some of the observational features of L1551
		such as the bending of the secondary jet.  }
   { 
			While the effects of orbital motion are negligible over the jets
			dynamical timeline, 
			their interaction has significant impact on their morphology.
			If the jets are not strictly parallel, as in most observed cases, we
			show that
			the magnetic field can help the collimation and refocusing of both of
			the two jets.  }

   \keywords{
		ISM: Herbig-Haro objects - ISM: individual objects: LDN 1551 IRS 5 - ISM: jets and outflows - stars: formation
               }

   \maketitle
%

\section{Introduction}

	
Well-collimated jets occur across a broad range of mass scales from young stellar objects to active galactic nuclei.
In young stellar objects (YSOs), jets and outflows are 
more easily observed at optical, infrared and UV wavelengths and the properties
are relatively well known \citep{2001ARA&A..39..403R,2007prpl.conf..231R}.
YSO jets are believed to be launched by means of a magnetic field which is 
anchored or frozen into a circumstellar disk, and pinched and twisted by the disks rotation. 
Once the pinched-in field reaches a critical angle, fast-rotating gas from the disk may be loaded onto the field lines and thus launched into the parent cloud in an outflow \citep{1982MNRAS.199..883B}.
The most accepted schools of thought diverge on whether the wind is launched primarily
from the inner disk region \citep[disk wind, ][]{1997A&A...319..340F,2000prpl.conf..759K} or the X-annulus where the young star's magnetosphere interacts with the disk (X-wind, \citep{2000prpl.conf..789S}). 
However in either case the jet is believed to brake the rotation of the disk 
and is thus essential to the star formation process, 
as it allows the star to accrete up to its final mass.

In all such models a single star with its associated circumstellar disk produces a bipolar outflow or jet. Estimates derived from studies of stellar populations, 
however, show that large numbers of 
binary and multiple star systems are expected
\citep{2002ApJ...581..654P,1995ApJ...443..625S,1993AJ....106.2005G,1991A&A...248..485D}. Do such systems present associated disks and jets ? 
It has been argued that the disk configuration in a binary system 
may either be in circumstellar 
disks or circumbinary disks depending on the star separation distance
\citep{2000prpl.conf..841H}.
On the other hand, it is well established
that multiple sources can be the 
source of HH objects, e.g. T Tau, IRAS 04325-1419, Z Cma, Sz 68, 
SR 24, S Cr A, AS 353 \citep{1993ApJ...408L..49R}.
Now, a multiple system may produce a single jet or outflow 
or a set of multiple outflows.
Although there should be ``many'' visible binary jets, only a 
few binary jets from binary protostars have been observed.
The examples known to the authors are compiled in Table 
\ref{BinaryOutflowTable}.
The frequency of existing binary jets is low compared to the 
large number of protostellar binary sources.

\begin{table*}
\begin{center}
\caption{Evidence of Binary Jets and Outflows?}
\label{BinaryOutflowTable}
\begin{tabular}{l l l l l}
\hline
\hline
Outflow (Source) & Natal Cloud & RA (J2000) & Dec (J2000) & Reference \\
\hline
HH154 (L1551 IRS 5) & Tau-Aur & 04 31 34.20 & +18 08 04.8 & \citet{2005ApJ...L}\\
HH1-2 - HH144 & Orion & 05 36 22.85 & -06 46 06.6 & \citet{1993ApJ...408L..49R} \\
HH111 - HH121 & Orion & 05 51 46.07 & +02 48 30.6 & \citet{1994AA...289L..19G} \\
L723 (IRAS 19156+1906) & Cep & 19 17 53.16 & +19 12 16.6 & \citet{2004RMxAC..21..100A} \\
HH288 (IRAS 00342+6347) & Cep& 00 37 11.07 & +64 03 59.8 & \citet{2001AA...375.1018G} \\
HH377 (IRAS 23011+6126) & Cep E & 23 03 13.9 & +61 42 21 & \citet{1997ApJ...474..749L} \\
IRAS 16293-2422 & $\rho$ Oph E & 16 32 22.8 & -24 28 33 D & \citet{2001ApJ...547..899H} \\
IRAS 20050+2720 & Cyg Rift & 20 07 06.7 & +27 28 53 & \citet{1995ApJ...445L..51B} \\
\hline
\end{tabular}
\end{center}
\end{table*}

Observations have also shown that multiple or quadrupolar jets may occur.
\citet{1993ApJ...408L..49R} discovered a second flow HH144 from the same source as HH1-2. 
\citet{1994A&A...289L..19G} observed a second flow (HH121) from the same source as the well-known HH111 outflow.
In both cases there are large angles of separation between the two jets.
This can be explained if the disks 
are not coplanar \citep{1994ARA&A..32..465M}.
However, observations of apparent double jets can also be explained by other means: \citet{1990ApJ...357..524A} imaged the outflow in L723 and found a distinct multi-polar morphology. They concluded that the lobes of the jets were in fact shell walls of two cavities swept clear by a single bipolar outflow. \citet{1991ApJ...376..615A} discovered a double radio source at the centre of the outflow structure and this led to a reappraisal of the of the double outflow theory \citep{1996ApJ...473L.123A,1997ApJ...489..734G,1998ASPC..132..303A,1998ApJ...504..334H,1999A&A...346..233P,2002ApJ...575..337S,2003RMxAC..15..135E,2004RMxAC..21..100A}.
Other example of multiple molecular outflows include IRAS 16293-2422, IRAS 20050+2720.

In order to explore these possibilities, and to investigate about the 
nature of jets from multiple systems, 
we perform numerical simulations of binary jets with typical physical quantities.
Our aim is to demonstrate the effects of their relative sizes and 
speeds on their interaction, in order to test if the jet evolution 
can lead to situations compatible with the few
binary jets observed, and with the corresponding observed properties.

As a starting point, we focus on jets emanated independently 
by two sources that launch the two outflows
with a small angle of separation.
A suitable test case for this scenario is the 
the binary jet HH 154, emanating from the source 
L1551 IRS 5.

Numerical models of optical jets have evolved over the past twenty years, from the original models by \citet{1981ApJ...247...52N} of the \citet{1974MNRAS.169..395B} twin-exhaust model,
to later models by
\citet{1984ApJ...276..560H,1987ApJ...316..323H,1988ApJ...326..323R} which were
1.5D bow shock models which explicitly tried to model observed features, to
axisymmetric and 3D geometries and including such physics as atomic radiative
cooling \citep{1990ApJ...360..370B,1998AJ....116.2943R,2000ApJ...540..192S,2000A&A...364..763R,2001A&A...367..959R}.

In our case we model a fully three-dimensional binary jet system using ATLAS, a new shock-capturing, multi-dimensional, constrained transport, adaptive-grid, directionally unsplit, higher-order Godunov astrophysical MHD code, ATLAS and examine its morphology and propagation dynamics.

The paper is structured as follows: in Section \ref{Observation_Section}
we describe the observed properties of the binary jet L1551 IRS 5, that 
constitutes our test case. 
In Section
\ref{Simulation_And_Code} we describe the methodology, the set of
equations used, the microphysics included, the numerical code 
and the initial and boundary conditions on the computational grid. 
In Sect.
\ref{Results} we present the results of the simulation, 
and 
finally in Sect. 5 we discuss our results and the insights brought forward by 
our model.

\section{Observations of jets from the binary Protostar L1551 IRS 5}
\label{Observation_Section}
 
 The picture being built up over 25 years of observations 
of the object L1551 IRS 5 and its 
associated jets and outflows
is that of a pair of YSOs each with its own associated disk and outflow, 
the whole structure embedded in a larger disk.

\cite{1976AJ.....81..638S} observed the near-infrared source IRS 5 within the Lynds molecular cloud 1551 (L1551).
The source has an infrared emission 
nebulosity \citep{1994ApJS...94..615H}, consistent with the presence 
of a channel perpendicular to the high-density disk, from which the light
from the central star escapes and irradiates the nebulosity.
\citet{1998Natur.395..355R} confirmed that L1551 IRS 5 was a binary system and show the first images of the circumbinary disks and the red lobes of the two jets.
Additionally they suggested that there is a circumbinary structure and a large-scale envelope around L1551 IRS 5.
\citet{1994ApJ...434L..75L} observed a large circumbinary disk of $\sim 160~\mathrm{AU}$ in diameter.
\citet{2003ApJ...586L.137R} confirmed the jet binarity at radio
wavelengths, while
\citet{2005ApJ...L} provided a value for binary separation of 40 AU. 
We add that recent observational evidence has suggested that 
the source may have a small
third companion \citep{2006ApJ...653..425L}.

\cite{1979AJ.....84..548C} discerned two fast Herbig-Haro objects near the IRS 5 source.
\cite{1980ApJ...239L..17S} saw for the first time the ``remarkable double-lobed
structure'' in L1551.
This was the among the first molecular bipolar outflows discovered.
\citet{1985ApJ...289L...5B} identified IRS 5 as a binary source.
\citet{1991ApJ...373L..23M} and \citet{1991ApJ...383..705P} identified a second outflow from the same source.
\citet{1991A&A...252..740M} also observed two independent 
rows of knots although these were interpreted as edges of a 
limb-brightened cavity, and were only identified as jets by \citet{1998ApJ...499L..75F} (hereafter FL), who 
measured a jet angular separation of approximately $20^{\circ}$.

Recent observations carried out with the Hubble Space Telescope and SUBARU
have further clarified the morphology of the flow \citep{2005A&A...436..983F}.
The optical jets extend southwest 
and disappear at approximately 1400AU from the IRS5. 
The north and south jets appear to be launched from the south and 
north disks respectively.
According to \citet{2005ApJ...L} the northern jet is faster 
with a radial velocity projected in the plane of the sky of $\sim$ 430 km~s$^{-1}$ whereas the 
southern jet has a radial velocity at most 65 km~s$^{-1}$.
\citet{2000PASJ...52...81I} observed a twisted morphology in 
the light of iron lines,
and argued that the precession of the sources 
is too slow to be responsible for this effect.
Possible other mechanisms can be magnetic in nature e.g. 
the Lorentz forces \citep{1998A&A...334..750F,2000IAUS..200P.112F} or the bending may be caused by the decrease in kinetic energy allowing the ambient magnetic pressure to be comparable to the ram pressure of the jet. 
Indeed, \citet{1988MNRAS.231P..39S} 
observed a high degree of polarization in the optical light 
emitted by the region
and concluded that it could be explained by a
toroidal field in the cloud around the outflows.
\citet{1997MNRAS.286..895L} also observed this ``peculiar pattern of
alignment''. The physical cause for the polarimetry pattern 
remains unexplained.

For completeness we mention also that 
X-ray emission from the region of the head of the north jet was observed by \citet{2002A&A...386..204F}.
\citet{2003ApJ...584..843B} performed a higher 
angular resolution study, finding
a source of X-rays in L1551 IRS 5 at the location of the source
which they attributed to either fast shocks or
reflected x-rays from IRS 5 scattered out through the outflow cavity.
\citet{2003ApJ...584..843B} put forward a number of models to explain the X-ray
emission, including the intriguing possiblity of a quasi-stationary X-ray
luminous shock maintained by interacting colliding winds 
between the two protostars.

\DeclareRobustCommand\itshape

\begin{table}[tb]
\caption{Observational Data for L1551 IRS 5 outflows}
\centering
\begin{minipage}{0.5\textwidth}
\begin{tabular}{l l}
\hline
\hline
N Jet Radial Velocity & 200-250 km~s$^{-1}$ \footnote{ \citet{2000AJ....119.1872H}} \\
S Jet Radial Velocity & 65 km~s$^{-1}$  \footnote{  \citet{2000AJ....119.1872H} }\\
Binary Separation & 45 AU \footnote{\citet{2005ApJ...L} } \\
Angle Between Jets & $20^{\circ}$  \footnote{\citet{1998ApJ...499L..75F}} \\
Orbital Period & 255 years \footnote{\citet{2000PASJ...52...81I}}\\
Ambient Density & 5000 cm$^{-3}$\footnote{\citet{1998ApJ...499L..75F}} \\
Jet Density & 500 cm$^{-3}$ \footnote{\citet{1998ApJ...499L..75F}} \\
Jet Length & 1400 AU \footnote{\citet{1998ApJ...499L..75F}}\\
\hline
\end{tabular}
\end{minipage}
\end{table}

\begin{figure}[ht]
\centering
\includegraphics[width=8cm]{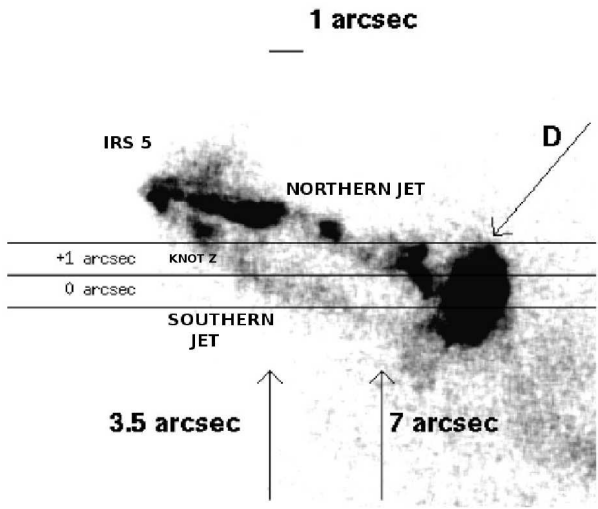}
\caption{
\citet{1998ApJ...499L..75F} provide a HST R-band image of the two jets from the
binary protostar L1551 IRS 5, located in the constellation Taurus. (Image courtesy of \citet{1998ApJ...499L..75F}) 
}
\label{fig:4-1} 
\end{figure}

\section{Methodology}\label{Simulation_And_Code}
Our main goal is to model binary jets. 
Contrary to single jet propagation, 
that can be modelled also analitically in an 
axisymmetric approximation, the propagation and interaction
of binary jets is an intrinsically three-dimensional problem.
A numerical study then becomes necessary, 
and requires the use of a fully 
three-dimensional time-dependent model with magnetic field and 
radiative cooling.
Constraints on current computational capacity require us to make 
a number of approximations however, 
that lead to the adoption of an ideal
magneto-hydrodynamics (MHD) approach. 
In this way we are able to include the 
effects of the magnetic field without losing the 
advantages of using a simple macroscopic 
description for a fluid particle.
We additionally include the non-ideal physics of optically thin atomic 
radiative cooling losses. 
Moreover, in our scheme 
the ionisation fraction of hydrogen is followed in the simulation.
This is important as several studies
\citep{ 1999A&A...342..717B,2006A&A...456..189P}
have demonstrated that this quantity in jets decouples
from the thermal gas conditions and follows the evolution of 
gas recombination.

\subsection{System of Equations}
The ideal MHD equations evolve in time 
the four quantities (two vector and two
scalar) which are conserved in a volume:
density $\rho$, momentum, $\rho \mathbf u$, 
magnetic flux density $\mathbf B$ and energy density, $E$.

\begin{equation}
\frac{\partial \rho}{\partial t}+\nabla\cdot(\rho \mathbf u)=0
\end{equation}

\begin{equation}
\frac{\partial}{\partial t}\left(
\rho \mathbf u
\right)
+\nabla
\cdot
\left[
\rho \mathbf u \otimes \mathbf u
+\left(
p^*
\right)
\mathbf {{\overline {\overline I}}}
+\mathbf{B} \otimes \mathbf{B}
\right]
=0
\end{equation}

\begin{equation}
\frac{\partial \mathbf{B} }{\partial t}+
\nabla
\cdot
(
\mathbf{u} \otimes \mathbf{B}-
\mathbf{B} \otimes \mathbf{u})
= 0
\end{equation}

\begin{equation}
\frac{\partial E}{\partial t}
+\nabla
\cdot
\left[
\left( E + p^*\right) \mathbf{u}
- (\mathbf{u}\cdot\mathbf{B})\mathbf{B}
\right] + L_{cooling} = 0
\end{equation}

where $E$ is the total energy, 
arising from the sum of three terms, namely kinetic, 
internal and magnetic energy:
\begin{equation} 
E =
\frac{1}{2}\rho |\mathbf{u}|^2 + 
\frac{p}{\gamma-1}+
\frac{1}{2} |\mathbf{B}|^2
\end{equation}
and $p^*$ is the total (thermal plus magnetic) pressure:
\begin{equation} 
p^* = p + \frac{1}{2}|\mathbf{B}|^2
\end{equation}

$L_{cooling}$ represents the losses due to 
optically thin radiative cooling, 
and is described in the next paragraph.
The units are chosen so that $\mathbf B$ absorbs a factor of $1/\sqrt {4 \pi}$.
The adiabatic index is $\gamma = 5/3$ for a monatomic gas throughout the simulations.
The equation of state is the ideal gas equation (p=nkT).

\subsection{Microphysics}

\begin{figure}[ht]
\centering
\includegraphics[width=8cm]{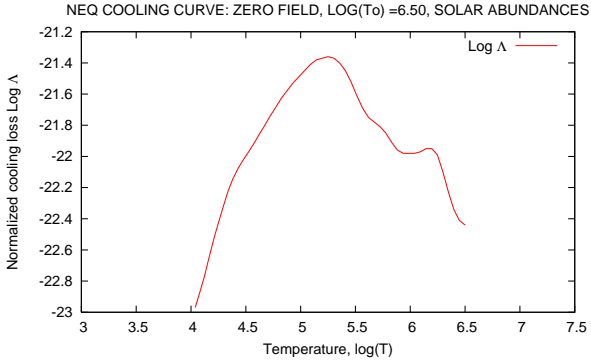}
\caption{
Log normalised radiative cooling loss plotted against log temperature \citep{1993ApJS...88..253S}.
}
\label{fig:4-33}\end{figure}

Protostellar jets are strongly cooled by
radiation of collisionally excited lines,
many of which are optically thin.
It it this emission which render the objects visible, 
particularly in H$\alpha$ and in the light of forbidden 
lines like the doublet of singly ionised sulphur at about 6700 \AA
([SII]$\lambda\lambda6716,6731$).
Modelling electron transitions directly, however, 
is neither feasible nor desirable in the fluid approximation.
Therefore, we 
represent the microphysics of energy lost due to radiative cooling using a
cooling function adapted from \citet{1993ApJS...88..253S} and depicted 
in Figure
\ref{fig:4-33}, which represents on a macroscopic scale 
the rate of energy loss at our temperature range for a 
sample (solar) set of abundances.
\citet{1993ApJS...88..253S} include the effects of electron 
collisional ionisation, radiative and dielectronic 
recombination and line radiation within their cooling function.

To calculate the energy loss in the gas one also needs to know 
the free electron density, which is 
computed explicitly at each 
time step and location in the fluid, as in \citet{1995MNRAS.272..785F}.
As mentioned above, this is necessary because the
ionisation fraction of hydrogen is not in equilibrium 
with the thermal properties of the gas, due to the fact 
that at the observed densities the recombination
times are long with respect to the crossing time of the visible 
portion of the jet. For a thorough discussion of this point 
see \citet{1997ApJS..109..517R, 1999A&A...342..717B}.

The cooling function shown in Figure \ref{fig:4-33} is highly non-linear.
 If we use the cooling timestep to evolve the overall jet the process will be
 extremely time-consuming.
 Therefore we separate the cooling from the dynamical evolution at each timestep,
 first calculating the dynamical change and then computing the radiative cooling using a short substep.

\subsection{The numerical code}

We computed the three-dimensional simulations using ATLAS, a new modular, parallel, shock-capturing, directionally unsplit, adaptive-grid, multi-dimensional, constrained transport, higher-order Godunov astrophysical
MHD code. Validation and verification tests, including but not limited to the
Orszag-Tang MHD vortex, the Brio-Wu and Ryu-Jones suite of shock tube
tests and 2D and 3D blast wave tests, have been run against the code to build confidence in its ability to form
correct solutions. ATLAS uses the PARAMESH \citep{par:00} hierarchical
block-structured adaptive mesh refinement for high effective resolution in areas
of physical interest. 
The solenoidal constraint (${\mathbf\nabla\cdot \mathbf{B}} =0$) is
preserved using a staggered mesh algorithm based on the \citet{307327} field transport method. 
ATLAS uses a Piecewise Parabolic scheme \citep{1984jcp_colella_woodward} to reconstruct the values at cell interfaces and the Roe-Balsara approximate MHD Riemann solver \citep{Roe81:_approx,Roe:1996:NEM} to explicitly compute the cell interfaces fluxes.
The multidimensional correction used to compute the transverse fluxes is the Corner Transport Upwind scheme of \citet{Colella90:_multida} as modified by \citet{Saltzman94:_an_uns}.
In our own case, the simulations were carried out using ATLAS on a 64 node cluster of the so-called ``Beowulf'' type.

\subsection{Initial and boundary conditions}

Consistently with the location of the object in Taurus, we 
assumed a distance of 140 parsecs to the jets. We used values derived from the
observations for the density, temperature and velocities, as described in Table 2.
The ambient medium is modelled
with a uniform density ($n_a=5 \times 10^{3}$~cm$^{-3}$) and temperature
($10^2 K$) and the jets are modelled with density $n_j=0.1 n_a$,
temperature 10$^4$K and velocities of 100 and 300~km~s$^{-1}$ respectively
\citep{2005ApJ...L}. Regarding the flow velocity, 
we take the radial velocities of \citet{2000AJ....119.1872H} 200~km~s$^{-1}$ and
60~km~s$^{-1}$ and deproject
them from the plane of the sky assuming (with \citet{2000AJ....119.1872H}) an
inclination angle of 45 degrees. 
In order to reproduce the knotty structure of the flows we also 
use a sinusoidally varying injection velocity
\citep{1990ApJ...364..601R}, assuming an amplitude of
$\pm$30\% in the velocity with a period of 8 years for each jet. 
We also stagger the launching of the two jets, arguing that the faster 
northern jet is launched 150 years after the slower southern jet.
The calculations of the launch times are based on estimating the age of the jets
from their lengths and current velocities, assuming the velocities have not
changed over time.
The velocity profile across the jet section is a positive cosine - with its maximum at v$_{jet}$ and its minimum (zero) 
at r$_{jet}$. This profile is chosen based on the high angular resolution observations 
of the DG Tau jet by \citet{2000ApJ...537L..49B}, where the highest velocities 
appear to be found in the axial region of the jet.

Typical postshock cooling lengths estimated using the plane parallel shock models of \citet{1987ApJ...316..323H} are of the order of 30 AU in the colliding outer wings of the binary jet bow shocks and in the internal shocks so will be
resolved in $\sim$12 points.

The initial jets' diameters are approximately 13 points each and the jet cross-sectional areas are resolved in approximately 150 square cells for each jet.

\begin{table}
\begin{center}
\begin{tabular}{l r}
\hline
\hline
Domain & $-840 $AU$<y,z<840$ AU \\
 & 0$<x<$1680 AU \\
Refinement & 5 levels \\
{ Max Resolution} & { 2.625 AU} \\
N Jet deprojected velocity & $v_j=300~$km~s$^{-1}$ \\
S Jet deprojected velocity & $v_j=100~$km~s$^{-1}$ \\
{ Jet radius} & $r_j=10^{14}$cm \\
Jet density & $n_j=500~$cm$^{-3}$ \\
Ambient density & $n_a=5000~$cm$^{-3}$ \\
Jet Temperature & $T_j=10^4 K$ \\
Ambient Temp & $T_a=100 K$ \\
{ Angle between jets} & {0 (Case I \& III)} \\
{ Angle between jets} & {10 (Case II)} \\
\hline
\end{tabular}
\caption{Initial conditions for simulation}
\end{center}
\end{table}

The boundary conditions are inflow for $r<r_j$, reflecting for $r>r_j$ along the inner x-boundary (x=0) and outflow along all other boundaries. 
\section{Results}\label{Results}

We present here results from fully 3D HD and MHD simulations of the 
jets. 

\subsection{Case I: Hydrodynamic case }

By using a simple hydrodynamical (HD) model, we were able to reproduce 
the twisted morphology of the L1551 IRS 5 outflow, that in this case
appears to arise from the interaction of the two jets. 
In the observations of L1551 IRS 5, the two jets emerge from a binary source
separated by 45 AU. 
The observed angular separation is 20 degrees which corresponds to a deprojected
angle of approximately 10 degrees. 
Initially, for the hydrodynamic case we assume the jets are parallel.
(In the next section (\ref{MHDjet}) we remove this simplifying assumption.)
We model the jet propagation and show that the interaction affects the propagation.

\begin{figure}[ht]
\centering
\includegraphics[width=8cm]{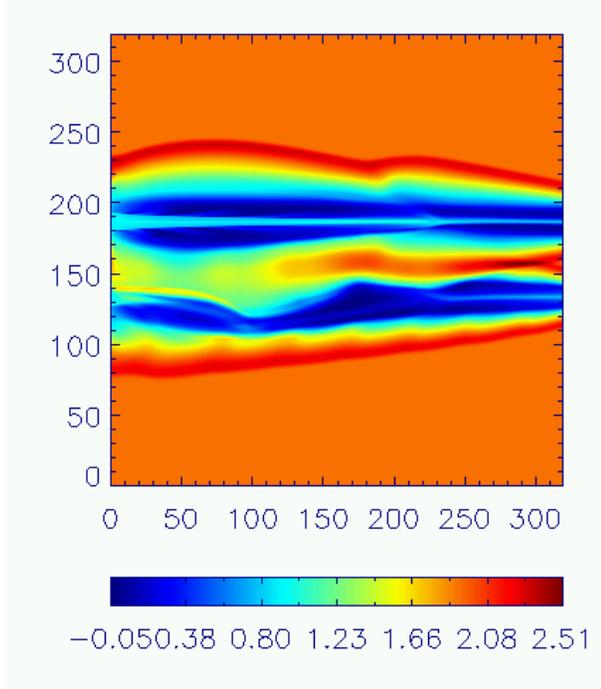}
\caption{ 
Midplane slice colour density map of 3-D binary jet simulation at $t=190~\mathrm{years}$.
The scale of the grid is $1680 \mathrm{AU}$ by $1680 \mathrm{AU}$.
}
\label{fig:4-5} 
\end{figure}

The density slice from the 3D HD simulation (see Figure \ref{fig:4-5}) 
clearly shows that the slow, secondary jet appears to bend close to the inlet.
Such a bend or kink is also observed in the slow, southern jet of 
L1551 IRS 5 about 4$^{\prime\prime}$ (560 AU at a distance of 140 pc
from the source \citep{2000PASJ...52...81I}).
In the simulation shown in Figure \ref{fig:4-5} a kink in the slow southern jet
is also visible - both in the density midplane cut shown in Figure \ref{fig:4-5}
and in the derived emission maps shown in \ref{fig:4-21}. This is caused by 
the bow shock of the fast jet interacting with the beam of the slow jet.
The northern jet has a Mach number $\sim 3$ times higher
than the slow southern jet and simply pushes it out of the way. This
reproduces the observed kink at 4$^{\prime\prime}$ - without the need for magnetic
fields. There is no noticeable reaction by the fast jet - possibly
indicating that the estimated velocity is too high.

\subsection{Case II: Magneto-hydrodynamic case}
\label{MHDjet}

In the second case, we consider the inclusion of the magnetic field 
and run a full magneto-hydrodynamic (MHD) simulation, trying
to reproduce the jet behaviour closer to the source.
The magnetic field provides a mechanism for the southern jet to change its
direction - hence we allow the jets to initially diverge.
The projected angle between the two jets is about 20 degrees
\citep{1998ApJ...499L..75F}
we use a deprojected angle of 10$^\circ$.
To 
redirect the two jets would require either a 
density contrast e.g. the wall of a conveniently shaped cavity 
or a magnetic hoop stress \citep{2003ApJ...584..843B}.

We use the magnetic field configuration based on the observations 
of \citet{1988MNRAS.231P..39S}, 
 that suggest the presence of a toroidal magnetic field around 
the system. The toroidal field may be
produced by the circumbinary disk twisting 
the frozen-in local magnetic field lines \citep[see, e.g.][ and the references
therein.]{2004ApJ...607L..43M}

We assume that the ambient medium is permeated by a toroidal field 
with a maximum value of B=10 $\mu$G. The magnetic
vector potential ${\mathbf A}$, where 
${\mathbf B = \nabla} \times {\mathbf A}$,
had the analytical form A$_y$= cos 2$\pi$ x + cos 2$\pi$ y.
This is an approximation to the real field which would be helical according to
the disk-twisting theory.

\begin{figure}[t]
\centering
\includegraphics[width=8cm]{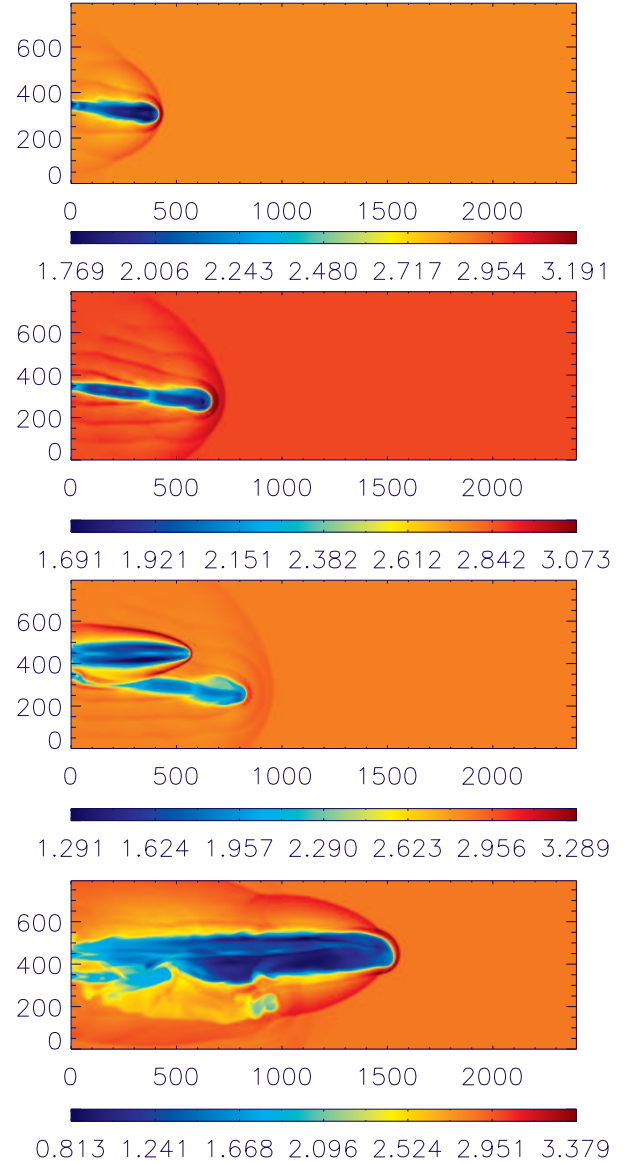}
\caption{ 
Time evolution of log jet total density 
in the MHD binary jets at t=42.5,85,127.5,170 years. 
The images shown are
2D midplane cuts through the 3D grid.
}
\label{fig:4-49}
\end{figure}

\begin{figure}[t]
\centering
\includegraphics[width=8cm]{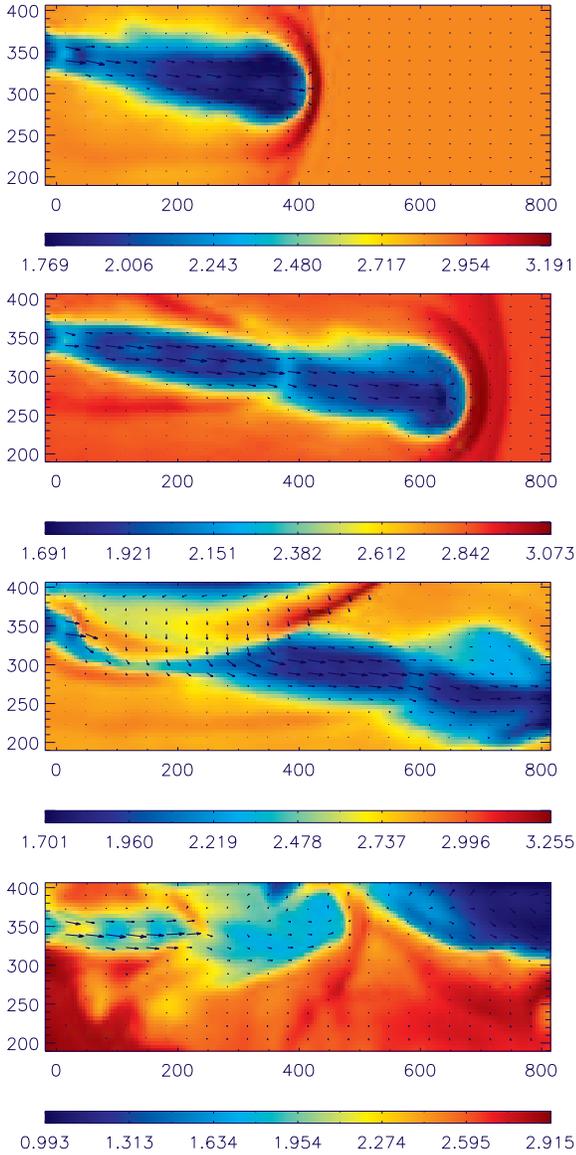}
\caption{ 
Section of Figure \ref{fig:4-49} showing velocity vectors.
}
\label{fig:4-50}
\end{figure}

Figure \ref{fig:4-49} shows a series of density slices from the 3D MHD
simulation. The binary outflow is modelled in full 3D with toroidal magnetic field.
Compared to the previous HD simulation, 
we have set the angle between the two jets to 
be close to 10$^\circ$. 
In the hydrodynamic case there is no mechanism to to force the jets to become
parallel. In the MHD case the Lorentz force can redirect both of the jets.
As a consequence, we would expect to see the two jets
propagating into two different directions on large scales. On the contrary,
the smaller jet is refocused along the faster jet. 
The secondary jet (i.e., the south low velocity one), bends towards
 the z-axis, impelled by Lorentz force. 
 This can be seen very clearly in Figure \ref{fig:4-49}, panel 1 for instance,
 where the slow jet is already moving towards the centre before the approach of
 the fast jet.
Eventually, the jet material from the slower 
jet gets merged into the faster one.

The observations of L1551 show both jets diverging from the source 
until about 4$^{\prime\prime}$ from the source when they change 
direction to pursue a roughly parallel course. The change in direction 
is most pronounced for the southern jet
which has the lower ram pressure and thus is easier to redirect.
In the simulation over the time evolution we see the southern jet change its
direction to move parallel to the axis of the toroidal field.
Hence, it appears that the hoop stress from the jet toroidal field slowly 
changes the direction of the jets.
Thus, magnetised binary jets will tend to collimate and refocus
along the direction of the fastest or strongest jet. 

\subsection{Case III: Orbiting binary jet}

\begin{figure}[ht]
\centering
\includegraphics[width=8cm]{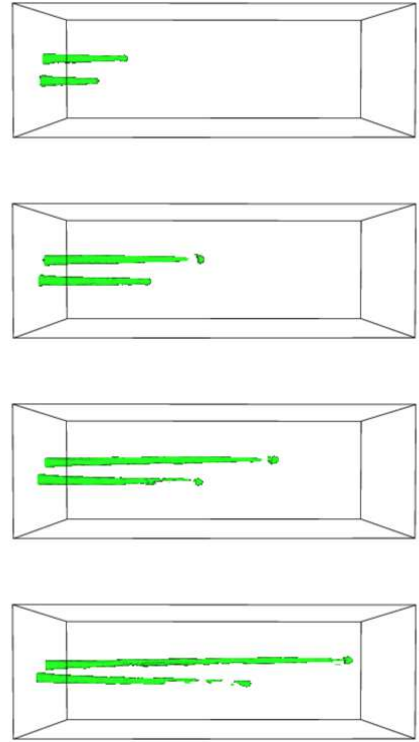}
\caption{ 
Surface plot of density of 3-D binary 
jet simulation at $t=5,10,40,60,90~\mathrm{years}$.
The orbital period of the source is 260 years.
The jet pair has performed 1/3 of a rotation.
}
\label{fig:4-6}
\end{figure}

We now explore the effects of orbital motion of the jet sources 
on the survival of the binary jets.
\citet{2002ApJ...568..733M} have modelled jets from orbiting sources
However they modelled just a single jet orbiting a central body and compared its
evolution to an analytical prediction.
Figure \ref{fig:4-6} shows a model of a binary jet from 
{ two sources orbiting} a common centre of mass. 
The jet crosses the grid in 90 years.
In this case, for 
simplicity we assume that the jets { are launched at the same time and their
velocities are 300 and 100 km~s$^{-1}$}.

We find that, by calculating the dynamical age of the jets 
in L1551 IRS 5, the orbital period 
of the binary $\sim$260 years is too long to have much effect on the fast 
northern jet (crossing time $\sim$90 years).

\subsection{Emission maps}
In order to properly compare the results of our simulations with 
the existing observations we have computed synthetic emission maps in the 
light of the forbidden lines of neutral oxygen
([OI]$\lambda$6300). In order to predict the 
line emission produced by the two jets in our model 
we use the density, temperature and the fraction
of ionised hydrogen as computed
by the numerical simulation, to find the emissivity 
 in each point of the jet gas. 
The emissivity $\epsilon$ 
is defined as the energy emitted at the wavelength $\lambda$ of 
the transition by a unit volume of gas per unit time.
Following \citet{2002RMxAC..13....8B}, the emissivity 
 can be expressed as:
\begin{equation}
\epsilon_{Z^i, \lambda} = A_{\lambda} \frac{hc}{\lambda} x_e n_H^2 
\left( \frac{Z^i}{Z}\right) 
\left( \frac{Z}{H}\right) 
\left( \frac{n_{upper}}{n(Z^i)} \right)
\end{equation}
where
$A_{\lambda}$ is the transition probability,
$n_H$ is the hydrogen nuclear number density,
$x_e$ is the ionised fraction of hydrogen,
 $ \frac{Z^i}{Z} $ is the fraction of the element Z ionised at level i, 
$\frac{Z}{H}$ is the abundance of element Z with respect to hydrogen, 
and $ \frac{n_{upper}}{n(Z^i)}$ is the fractional population of the upper level
of the transition.
We assume solar abundances for the elements in the fluid.
{
For O,  following \citet{1999A&A...342..717B},
we can determine the fractional ionisation as a function of $x_e$
assuming the relationship 
\begin{equation}
\frac{O^+}{O^0}=\frac{(C_O + \delta_O )x_e}{(\alpha_O - \delta_O) x_e + \delta'_O} 
\end{equation}
holds, where $C_O$ is the collisional ionization rate, $\alpha_O$ the direct plus dielectronic recombination rates, $\delta_O$ and $\delta'_0$ are the direct and inverse charge ionisation exchange rates respectively.}
The fractional population of the upper level, 
$ \frac{n_{upper}}{n(Z^i)}$, is basically a function of {the} free electron density 
and {the} electron temperature, and is determined by a 5-level linear system of 
equations describing the statistical equilibrium of the level populations 
subject to collisional excitation and de-excitation, and to 
spontaneous radiative decay.
A more complete description of the method can be found in
\citet{1995A&A...296..185B, 1999A&A...342..717B} and 
\citet{2002RMxAC..13....8B}.

\begin{figure}[ht]
\centering
\includegraphics[width=8cm]{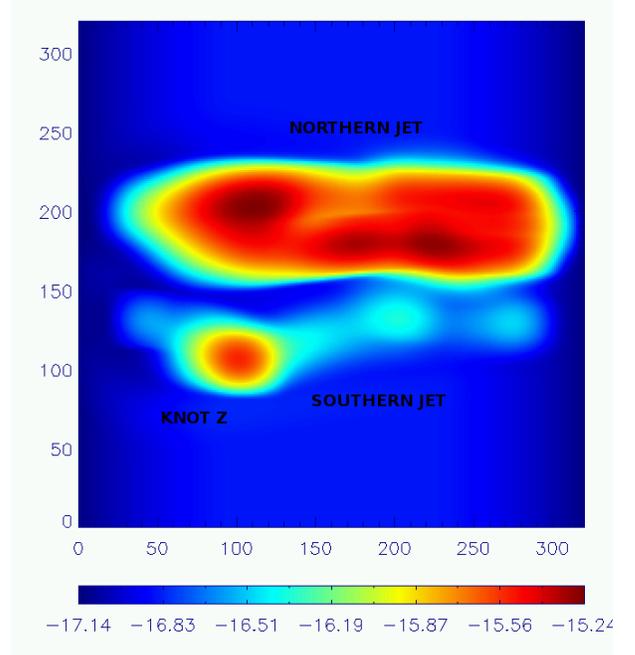}
\caption{  Synthetic [OI]$\lambda$6300 
emission map derived from the 
physical parameters of our simulation in the hydrodynamic case as depicted in
Figure \ref{fig:4-5}.
to be compared with corresponding HST observations of the 
L1551 IRS 5 jet (HH154) (See text)
 The image shows the northern and southern jets at a time t=190 years. 
 Knot Z indicates the point where the bow shock of the northern jet collides
 with the beam of the southern jet.
 The units are erg cm$^{-2}$ s$^{-1}$ arcsec$^{-1}$.
}
\label{fig:4-21} 
\end{figure}

The emissivity values calculated in the 3D space 
from {our simulation} are then integrated along {lines of sight}, assuming an inclination 
angle of 45 degrees. 
The result is then convolved with the beam of the 
instrument, which {corresponds to} an angular resolution of 0.$''$1 in the HST case. 
 
The synthetic emission map produced in this way is shown in 
Figure \ref{fig:4-21}. It is clear that the interaction has a 
strong observational signature. The second jet is virtually 
obliterated but remains visible at the point where the 
fast jet's bow shock impinges on its beam. 
The emission map is quite different to what is seen in the density 
slice. 
The knotty structure of the southern jet and the asymmetric structure of the
northern jet are only apparent in the emission map.
These may be compared with the observations in Figure \ref{fig:4-1}.
Even though the image has been convolved to account for the beam
resolution
a peak in emission is visible in the southern jet where the bow shock of the
fast jet is striking the beam.
In the observations a similar peak in the emission is visible near the source of
the southern jet (Knot Z in Figure \ref{fig:4-1}).

\section{Discussion}\label{Discussion and Summary}

	The L1551 IRS 5 jet pair can be seen as a pathological 
object unlike any other observed at such close range.
In fact many effects, like binary source orbiting, peculiar 
environment, density parameters, magnetic fields play a role 
in sculpting this most unusual object. To attempt an explanation of the 
observed phenomena that can take into account all the above mentioned elements,
a full 3D numerical treatment is necessary.
In this paper we have presented 3D HD and MHD simulations of 
interacting binary jets that may help explain the L 1551-IRS5 case. 
We have used the code ATLAS including radiative cooling 
and non-equilibrium ionisation treatment.
The main results can be summarised as follows: 

\begin{itemize}
\item We can reproduce the kink and bending of the secondary jet and the knotty
structure of the two components as observed for L1551 IRS 5.
\item Over a long time-scale, we see precession induced 
by the orbital motion of the source. 
On the short lifetime of the jets from L1551, this effect is negligible. 
\item If the jets are not strictly parallel, 
as in most observed cases, we show that
the magnetic field can help the collimation and refocusing of both the jets.
In fact the toroidal field affects the motion of the southern jet; and
this magnetically driven change in direction together with the interaction
of the bow shock of the fast jet with the beam of the slow jet 
contribute to the distinctive morphology.
\item We have produced emission maps in the [OI] lines
which can be compared with observations.
We show that the kink structure of interacting jets is 
still apparent in the synthetic 
observations.
\end{itemize}

Some aspects however remain to be clarified. For example, 
one of them is the nature of the the source of the X-ray emission.
\citet{2003ApJ...584..843B} find a source of X-rays in L1551 IRS 5 
which they attribute to either fast shocks or reflected x-rays 
scattered out through the outflow cavity.
The fast shocks may be caused by jet collimation 
which could be magnetic in nature. 
In our model there is a peak in emission (Knot Z in Figure \ref{fig:4-1}) which
appears to come from the colliding winds.
The Bally model suggests that the X-rays come from a moving source at the base
of the jets. If the density ratio is $n_{ambient}/n_{jet}=10$ then the shock speed
will be too slow to achieve X-ray emission temperatures. If on
the other hand the density ratio is lower i.e. 0.1 the jet is moving into a less
dense region swept out by the slower jet the shock velocities will be higher. To get up to the observed shock velocity the density contrast would
need to be much lower than observed e.g. 0.01 would give a velocity in the range required.
We also could assume that it is caused by magnetic reconnection in the
interval between the two jets where the field is compressed by the pair of bow
shocks. 

One other interesting aspect to be discussed is the
implications for jet launching of the binarity of the source. 
The binary provides a vastly different natal environment for a protostar and its
attendant jets and outflows than that envisaged in a single star formation.
We identify three main differences for jet launching for binaries, orbital motion, circumbinary material and interaction with a jet.
In this work we showed that the impact upon the jet beam of the bow shock of its
neighbour is large enough to be observable as a knot of emission moving along
along the jet. The orbital motion is of too long a period to affect the jet
propagation. The role of the circumbinary material may be to twist up the magnetic field into the toroidal shape which appears in the polarized light.
In conclusion, our simulations have demonstrated to be able to reproduce 
the observed signatures in this source and to give powerful constraints 
on the physical effects at play. 
A larger number of observed cases has however to be modeled with our code 
in order to give a complete view of binary jet production.

\section*{Acknowledgements}
This work was carried out as part of the CosmoGrid project, funded under the Programme for Research in Third Level Institutions (PRTLI) administered by the Irish Higher Education Authority under the National Development Plan and with partial support from the European Regional Development Fund. The PARAMESH software used in this work was developed at the NASA Goddard Space Flight Center under the HPCC and ESTO/CT projects. Some of this work was carried out using ICHEC/Cosmogrid computers and under the HPC-EUROPA project (RII3-CT-2003-506079),with the support of the European Community Research Infrastructure Action under the FP6 Structuring the European Research Area Programme).
The authors acknowledge support
through the Marie Curie Research Training Network JETSET (Jet Simulations,
Experiments and Theory) under contract MRTN-CT-2004-005592.

\bibliographystyle{aa}

\end{document}